\documentclass[twocolumn, 3p, number, sort&compress]{elsarticle}
\usepackage{relsize}
\usepackage{upgreek}

\usepackage{amsmath,amssymb,graphicx,hyperref}
\usepackage[utf8]{inputenc}
\usepackage[T1]{fontenc}
\usepackage{multirow}
\usepackage{fancyvrb}
\usepackage{mathtools}
\usepackage{color}
\usepackage[usenames,dvipsnames]{xcolor}
\usepackage{pdflscape,array,booktabs}

\usepackage{siunitx}
\usepackage{matlab-prettifier}
\usepackage{listings}
\usepackage{bm}
\newcommand{\pd}{\partial}
\def\d{\mathrm{d}}
\renewcommand{\b}{\boldsymbol}
\def\b#1{\mathbf{#1}}
\newcommand{\epsr}{\epsilon}
\newcommand{\Qnmp}{\partial Q_n^m}
\newcommand{\Pnmp}{\partial P_n^m}
\newcommand{\pot}{\Phi}

\title{Electrostatic limit of the $T$-matrix for electromagnetic scattering: Exact results for spheroidal particles}

\begin{document}
\date{\today}
\author{Matt R. A. Maji\'c}
\ead{mattmajic@gmail.com}
\author{Finnian Gray} 
\author{Baptiste Augui\'e} 
\ead{baptiste.auguie@gmail.com}
\author{Eric C. Le Ru\corref{cor1}}
\ead{eric.leru@vuw.ac.nz}

\address{The MacDiarmid Institute for Advanced Materials and Nanotechnology,
School of Chemical and Physical Sciences, Victoria University of Wellington,
PO Box 600, Wellington 6140, New Zealand}
\cortext[cor1]{Corresponding author}

\begin{abstract}
The $T$-matrix, often obtained with Waterman's extended boundary condition method (EBCM), is a widely-used tool for fast calculations of electromagnetic scattering by particles.
Here we investigate the quasistatic or long-wavelength limit of this approach, where it reduces to an electrostatics problem.
We first present a fully electrostatic version of the EBCM/T-matrix method (dubbed ES-EBCM). Explicit expressions are then given to quantitatively express the long-wavelength limit of the EBCM matrix elements in terms of those of the ES-EBCM formalism. 
From this connection we deduce a number of symmetry properties of the ES-EBCM matrices. 
We then investigate the matrix elements of the ES-EBCM formalism in the special case of prolate spheroids. Using the general electrostatic solution in spheroidal coordinates, we derive fully analytic expressions (in the form of finite sums) for all matrix elements.
Those can be used for example for studies of the convergence of the $T$-matrix formalism. We illustrate this by discussing
the validity of the Rayleigh hypothesis, where analytical expressions highlight clearly the link with analytical continuation of series.
\end{abstract}
\maketitle


\section{Introduction}

In electromagnetic scattering, it is often convenient to express the electric field in terms of
multipoles, e.g.
\begin{align}
\mathbf{E}(\mathbf{r}) = \sum_{n,m} a_n^m \mathbf{M}_{nm}(\mathbf{r}) + b_n^m \mathbf{N}_{nm}(\mathbf{r})
\end{align}
where $\mathbf{M}_{nm}$ and $\mathbf{N}_{nm}$ are the electric and magnetic multipolar fields,
also called vector spherical wavefunctions, and form a complete basis of divergence-less solutions of Helmoltz equation found using separation of variables in spherical coordinates \cite{1953Morse,1983Bohren,2002Mishchenko,2009Rother}.
The coefficients of the series expansion for the scattered field depend linearly on those
of the incident field, and this relation defines the transition $T$-matrix \cite{2002Mishchenko}.
Within the extended boundary condition method (EBCM) \cite{1990Barber,2002Mishchenko}, also called null-field method \cite{2006Doicu}
or $T$-matrix method, the $T$-matrix is computed from the division of two matrices whose matrix elements are given by surface integrals on the particle surface.
This approach was developed by Waterman in the 1960-70s \cite{1965Waterman,1971WatermanPRD,1975BarberAO} and remains one of the most powerful techniques for the study of electromagnetic scattering by particles.
It has been extensively studied and applied in many areas to compute the scattering properties of particles of arbitrary shapes \cite{2014Mishchenko} with a particular emphasis on axisymmetric particles.

Only a few studies have however studied its potential use to solve electrostatic problems involving dielectric particles. One approach is
to take the long-wavelength limit of the standard EBCM/$T$-matrix formalism. This was 
discussed by Waterman in the special two-dimensional case of infinite cylinders \cite{2007WatermanJOSAA}.
More recently, Farafonov and co-workers have developed an electrostatic (ES) equivalent of the $T$-matrix/EBCM formulation \cite{2014Farafonov,2016Farafonov}, hereafter referred to as ES-EBCM.
They then exploited it to derive asymptotic expansions (for large multipole orders) for the matrix elements arising from the EBCM formalism and to study the validity of the Rayleigh hypothesis for electromagnetic scattering by a particle. The hypothesis states that the scattered field expansion converges everywhere outside the particle \cite{2002Mishchenko,2009Rother,2016AuguieJO}. While it is generally accepted that the Rayleigh hypothesis is not always valid, its range of validity
and its link to the singularity of the solutions are still the subject of investigations \cite{2009Rother,2009WauerOC,2016AuguieJO,2016Farafonov}.

In this work we show that analytic expressions can in fact be obtained for all ES-EBCM matrix elements in the special case of spheroids. For this we use expansions of solid spherical harmonics in terms of solid spheroidal harmonics and vice-versa.
The analytic $T$-matrix extends the previously-found asymptotic behavior of the matrix elements. It moreover provides an analytic illustration of the region of validity of the Rayleigh hypothesis and its link to the analytical continuation of the series solution. 

The manuscript is organized as follows. In Sec.~\ref{SecESEBCM}, we re-derive a fully self-contained ES-EBCM formulation, and establish the link to the EBCM quantitatively. This allows us to discuss the symmetry properties of the ES-EBCM matrices from their EBCM counterparts.
We then focus on the case of spheroids, first using the ES-EBCM integrals (Secs.~\ref{SecSpheroids}). A simple expression is derived for the first column of the $T$-matrix, which allows us to discuss the Rayleigh hypothesis and its connection with the analytic continuation of the series of the scattered field.
In Sec.~\ref{SecAnalytic} we then use the separation of variable method in spheroidal coordinates, and relationships between spherical and spheroidal harmonics, to obtain analytic expressions for all matrix elements.

\section{The EBCM for the electrostatics problem}
\label{SecESEBCM}

As described in Refs. \cite{2014Farafonov,2016Farafonov}, the EBCM for electrostatics (ES-EBCM) can be
derived from scratch following the same approach as for electromagnetic scattering. For practical reasons, the basis functions are different to the long-wavelength limit of the spherical wave-functions, which makes quantitative connections between the two approaches cumbersome. Here we will therefore use different notations and conventions to previous work \cite{2014Farafonov,2016Farafonov} to simplify the link to the full wave EBCM. We summarize below the main definitions and derivations of the ES-EBCM and provide this quantitative link with the EBCM.

We consider a dielectric particle embedded in a medium with dielectric constants $\epsilon_2$ and $\epsilon_1$, respectively, and an external field $\mathbf{E}_\mathrm{ext}(\mathbf{r})$. The corresponding potential is $\pot_\mathrm{ext}(\mathbf{r})$, and we denote $\pot_\mathrm{out}(\b r)$ and $\pot_\mathrm{in}(\b r)$ the (total) potentials outside and inside the particle.

\subsection{Surface-integral formulation}

The electrostatics problem can be formulated in terms of surface integral equations \cite{2007VanBladel}.
For two points $\b{r}$ and $\b{r'}$, the free space Green's function is
\begin{align}
G(\b{r}, \b{r}')=&\frac{1}{4\pi|\b{r}-\b{r}'|}.
\end{align} 
Using Laplace's equation, $\nabla^2 \pot_\mathrm{in}=0$, and Green's second identity with $\pot_\mathrm{in}$ and $G$ on the interior volume, we obtain:
\begin{align}
\int_S \left[\pot_\mathrm{in}(\b{r}')\frac{\pd G}{\pd n'_-} \right.&\left. 
- G \frac{\pd \pot_\mathrm{in}}{\pd n'_-}(\b{r}')\right]dS' \nonumber\\
& = 
\begin{cases}
-\pot_\mathrm{in}(\b{r}) &\mbox{if } \b{r}\in V,\\
0 &\mbox{if } \b{r} \notin V.
\end{cases}
\label{EquSIE1}
\end{align}
where $\partial/\partial n'_-$ is the derivative just inside the surface, with respect to surface normal (and $\partial/\partial n'_+$ is the derivative just outside).

For the exterior volume $\bar{V}$, we note that $\pot_\mathrm{out}$ does not strictly satisfy Laplace's equation everywhere because of sources $\rho_\infty$ at infinity associated with the external potential, i.e. $\nabla^2 \pot_\mathrm{ext}=-\rho_\infty / \epsilon_0$ and $\pot_\mathrm{ext}=\int_{\bar{V}} G \rho_\infty / \epsilon_0 dV'$. Using Green's second identity with $\pot_\mathrm{out}$ and $G$ on the exterior volume, we obtain (note the change of sign to retain the normal pointing outward):
\begin{align}
\int_S  \left[\pot_\mathrm{out}(\b{r}')\frac{\pd G}{\pd n'_+} \right.&\left. - G  \frac{\pd \pot_\mathrm{out}}{\pd n'_+}(\b{r}')\right]  dS' \nonumber \\
& =  \begin{cases}
\pot_\mathrm{sca}(\b{r}) &\mbox{if } \b{r}\notin V,\\
-\pot_\mathrm{ext}(\b{r}) &\mbox{if } \b{r} \in V.
\end{cases}
\label{EquSIE2}
\end{align}
where $\pot_\mathrm{out}=\pot_\mathrm{ext} + \pot_\mathrm{sca}$, with $\pot_\mathrm{sca}$
the ``scattered'' potential. \footnote{The denomination ``scattered'' potential is aligned with the full-wave treatment, but the term ``reflected'' potential
would be more suited to a purely electrostatics problem.}
This expression is analogous to the null-field equations of the EBCM.

Using the boundary conditions on the surface: $\pot_\mathrm{out}=\pot_\mathrm{in}$ and $\epsilon_1\pd\pot_\mathrm{out} / \pd n_+=\epsilon_2\pd\pot_\mathrm{in} / \pd n_-$,
we can combine the above equations to eliminate the Green's function derivative (note that $\pd G / \pd n'_+=\pd G / \pd n'_-$) and get:
\begin{align}
\left(\epsr-1\right)\int_S \frac{\pd \pot_\mathrm{in}}{\pd n'_-}(\b{r}') & G(\b{r},\b{r}')dS' \nonumber \\
&= \begin{cases}
\pot_\mathrm{ext}-\pot_\mathrm{in}, &\mbox{if } \b{r}\in V\\
-\pot_\mathrm{sca}, &\mbox{if } \b{r} \notin V
\end{cases} \label{int_eqs}
\end{align}
where $\epsr=\epsilon_2 / \epsilon_1$ is the relative dielectric constant.

\subsection{Expansion in terms of solid spherical harmonics}

The next step is to expand Green's function and the potentials onto the basis of interest, here the standard solutions of Laplace's equations in spherical coordinates, chosen here as:
\begin{align}
\Psi^{(1)}_{nm}(\b{r})=& A_n^m \left(\frac{r}{R}\right)^n P_n^m(\cos \theta)e^{im\phi}, \\
\Psi^{(3)}_{nm}(\b{r})=& A_n^m \left(\frac{R}{r}\right)^{n+1} P_n^m(\cos \theta)e^{im\phi}, \label{Vexpnd}\\ 
A_n^m =& \frac{1}{\sqrt{4\pi}}\sqrt{\frac{(n-m)!}{(n+m)!}}. \label{Anm}
\end{align}
We include the arbitrary length $R$ so that the basis functions are dimensionless. Without it, the matrix elements would have dimensions dependent on the row and column, which is not ideal. The full-wave EBCM basis-functions are also commonly defined as dimensionless \cite{2002Mishchenko}. $R$ may be set as a characteristic length of the problem, although physical predictions are independent of the choice of basis functions and therefore independent of $R$. 

Our choice of normalization is slightly different to that used by Farafonov {\it et al.},
and will simplify the connection with the full-wave EBCM and with the approach using spheroidal coordinates in Sec.~\ref{SecAnalytic}.
These basis functions satisfy the following orthogonality relations on a sphere of radius $R$:
\begin{align}
\int_{r=R} \Psi^{(i)}_{nm}(\b{r}) \Psi^{(j)}_{n'm'}(\b{r})dS= \frac{\delta_{n,n'} \delta_{m,m'}}{2n+1}R^2,
\end{align}
and Green's function can be expanded as \cite{1998Jackson}:
\begin{align}
G(\b{r}, \b{r}')=\begin{dcases}\frac{1}{R}\sum_{\bar{n}}\left[\Psi_{\bar{n}}^{(1)}(\b{r}')\right]^*\Psi_{\bar{n}}^{(3)}(\b{r}), & \mbox{if } r>r' \\
\frac{1}{R}\sum_{\bar{n}}\left[\Psi_{\bar{n}}^{(3)}(\b{r}')\right]^*\Psi_{\bar{n}}^{(1)}(\b{r}), & \mbox{if }  r<r'\end{dcases}
\end{align}
where the combined index notation $\bar{n}$ is equivalent to both indices $n,m$ (with $|m|\le n$).

The external and internal potentials can be written as series in terms of $\Psi^{(1)}_{nm}$, which are regular at the origin, while the scattered potential is expanded in terms of $\Psi^{(3)}_{nm}$, which vanish at infinity. We here use notation in line with the electromagnetic treatment as presented in Ref. \cite{2002Mishchenko}, explicitly:
\begin{align}
\pot_\mathrm{ext}(\b{r})=&\sum_{n,m} ~\tilde{b}_n^m ~\Psi^{(1)}_{nm}(\b{r}), \label{EqnPhiExtSph}\\[0.2cm]
\pot_\mathrm{in}(\b{r})=&\sum_{n,m} ~\tilde{d}_n^m ~\Psi^{(1)}_{nm}(\b{r}),\\[0.2cm]
\pot_\mathrm{sca}(\b{r})=&\sum_{n,m} ~\tilde{q}_n^m ~\Psi^{(3)}_{nm}(\b{r}). \label{EqnPhiScaSph}
\end{align}

\subsection{Derivation of the ES-EBCM formula}

To solve the problem (determine $\pot_\mathrm{sca}$ and $\pot_\mathrm{in}$), we substitute the series expansions of the potentials and of Green's function
into the integral equations (Eqs. \ref{int_eqs}) and use the orthogonality of the basis functions to obtain
\begin{align}
\begin{array}{rl}
-\tilde{q}_{\bar{n}} =& (\epsr-1)\sum_{\bar{k}} ~\tilde{L}^{11}_{\bar{n}\bar{k}}~\tilde{d}_{\bar{k}}, \\[0.5cm]
\tilde{b}_{\bar{n}}-\tilde{d}_{\bar{n}} =& (\epsr-1)\sum_{\bar{k}} ~\tilde{L}^{31}_{\bar{n}\bar{k}}~\tilde{d}_{\bar{k}},
\end{array}
 \label{lin_eqs}
\end{align}
where (similarly to Ref. \cite{2014Farafonov}):
\begin{align}
\tilde{L}^{11}_{\bar{n}\bar{k}} = \frac{1}{R}\int_S \left[\Psi_{\bar{n}}^{(1)}(\b{r}')\right]^* \frac{\pd \Psi^{(1)}_{\bar{k}}}{\pd n'}(\b{r}')dS',\\[0.2cm]
\tilde{L}^{31}_{\bar{n}\bar{k}} = \frac{1}{R}\int_S \left[\Psi_{\bar{n}}^{(3)}(\b{r}')\right]^* \frac{\pd \Psi^{(1)}_{\bar{k}}}{\pd n'}(\b{r}')dS'.
\end{align}
The matrices $\b L^{11}$ and $\b L^{31}$ obtained by Farafonov \cite{2014Farafonov,2016Farafonov} differ from these by some normalization and dimensional coefficients due to different basis functions. 

Eqs. \ref{lin_eqs} form an infinite-dimensional linear system between the expansion coefficients, and just as with the EBCM, these can be written succinctly using matrix notations:
\begin{align}
\tilde{\b{q}}=-\tilde{\b{P}}\tilde{\b{d}},\qquad &\tilde{\b{b}}=\tilde{\b{Q}}\tilde{\b{d}},\qquad \tilde{\b{q}}= \tilde{\b{T}} \tilde{\b{b}},
\end{align}
where $\tilde{\b{b}}, \tilde{\b{d}}$, $\tilde{\b{q}}$ are vectors containing elements $\tilde{b}_{\bar{n}}$, $\tilde{d}_{\bar{n}}$, $\tilde{q}_{\bar{n}}$, and $\tilde{\b{P}}$, $\tilde{\b{Q}}$, $\tilde{\b{T}}$ are infinite-dimensional matrices given by:
\begin{align}
\tilde{\b P} &=  (\epsr -1)\tilde{\b L}^{11} \nonumber\\[0.2cm]
\tilde{\b Q} &=  {\b I} + (\epsr -1) \tilde{\b L}^{31} \nonumber\\[0.2cm]
\tilde{\b T} &= -\tilde{\b P}\tilde{\b Q}^{-1} \label{PQT_ES}
\end{align}

Physically, $\tilde{\b P}$ is the coupling matrix for the multipolar component of order $\bar{n}$ of the field created outside
by a multipolar field of order $\bar{k}$ inside.
$\tilde{\b Q}^{-1}$ is the coupling matrix for the multipolar component of order $\bar{n}$ of the field created inside
the particle when excited by an external multipolar field of order $\bar{k}$.
It is interesting to note that the matrices $\tilde{\b L}^{11}$ and $\tilde{\b L}^{31}$ are independent of permitivity and depend on shape only.
In fact, the matrix $\tilde{\b L}^{31}$ is a generalization of the depolarization factor introduced in the standard electrostatics solution of the ellipsoid \cite{1941Stratton}.

We also note that Green's second identity on $[\Psi_{\bar{n}}^{(1)}]^*$ and $\Psi^{(1)}_{\bar{k}}$
implies that the matrix $\tilde{\b L}^{11}$ is Hermitian, i.e. $\tilde{L}^{11}_{\bar{n}\bar{k}}=[\tilde{L}^{11}_{\bar{k}\bar{n}}]^*$.

\subsection{Link with the standard EBCM formulation}
\label{SecLimits}

The connection between the ES-EBCM and EBCM formulations was discussed in
Ref.~\cite{2013LeRuPRA}, but with slightly different definitions and only for the $T$-matrix, so for completeness we here adapt and expand the main results.
We use the standard EBCM notations of Ref.~\cite{2002Mishchenko} except for the $\b {RgQ}$ matrix denoted $\b P$ here.
The link can be established by considering the long-wavelength limit of the EBCM. 
We do not here consider the magnetostatic limit or the limit of the coupling between magnetic and electric multipoles,
so only the block ${\b T}^{22}$ relating to electric/electric multipole coupling needs to be considered.
In the long wavelength limit, the electric multipole fields reduce to \cite{2013LeRuPRA}:\footnote{Note that there is an error for ${\b N}^{(3)}_{nm}$ in Eq. (27) of Ref. \cite{2013LeRuPRA} as the exponent for $k$ not should be $n+1$, but $n+2$ (as corrected here).} 
\begin{align}
\mathbf{N}_{nm}^{(1)} &\rightarrow A_n^m B_n k^{n-1}~{\b \nabla}[r^nP_n^m(\cos\theta)e^{im\phi}], \label{limN1}\\ 
\vspace{.2cm}
\mathbf{N}_{nm}^{(3)} &\rightarrow\frac{A_n^m}{B_n}\frac{i}{k^{n+2}}\sqrt{\frac{n}{n+1}}~{\b \nabla} \frac{P_n^m(\cos\theta)e^{im\phi}}{r^{n+1}} \label{limN3}
\end{align}
where $\mathbf{N}_{nm}^{(1)}$ and $\mathbf{N}_{nm}^{(3)}$ are the regular and irregular vector spherical wavefunctions for electric multipoles, $k$ is the wavenumber ($k_2$ inside the particle, $k_1$ outside), and we have defined for convenience:
\begin{align}
B_n = \frac{1}{(2n-1)!!}\sqrt{\frac{(n+1)}{n(2n+1)}}.
\end{align}
Comparing the general expansions of the incident and scattered electric fields and potentials via $\mathbf{E}=-\nabla \pot$, and using \ref{limN1} and \ref{limN3} we derive:
\begin{align}
\tilde{b}_{\bar{n}}&=-R(k_1R)^{n-1} B_n ~b_{\bar{n}}, \\[0.2cm]
\tilde{d}_{\bar{n}}&=-Rs^{n-1} (k_1R)^{n-1}B_n ~d_{\bar{n}}, \\[0.2cm]
\tilde{q}_{\bar{n}}&=-\frac{iR}{(k_1R)^{n+2}B_n}~q_{\bar{n}},
\end{align}
where $s=k_2/k_1=sqrt{\epsr}$ is the relative refractive index of the particle, and $b_{\bar{n}}$, $d_{\bar{n}}$, $q_{\bar{n}}$ are the expansion coefficients of the electric multipole component of the incident, internal and scattered electric fields in the EBCM \cite{2002Mishchenko}.

From those, we deduce the relations between the matrices in the $\lambda\rightarrow \infty$ limit:
\begin{align}
T^{22}_{\bar{n}\bar{k}}(\lambda\rightarrow \infty) &= -i (k_1R)^{n+k+1}~B_n B_k  ~\tilde{T}_{\bar{n}\bar{k}}, \nonumber\\[0.2cm]
P^{22}_{\bar{n}\bar{k}}(\lambda\rightarrow \infty) &= -i s^{k-1} (k_1R)^{n+k+1}~B_n B_k  ~ \tilde{P}_{\bar{n}\bar{k}}, \nonumber\\[0.2cm]
Q^{22}_{\bar{n}\bar{k}}(\lambda\rightarrow \infty) &= s^{k-1} (k_1R)^{k-n}~\frac{B_k}{B_n}  ~ \tilde{Q}_{\bar{n}\bar{k}}.
\label{EBCMvsESBCM}
\end{align}
Again the choice of $R$ does not affect the long-wavelength limit of the EBCM. This can be checked from Eqs. \ref{Lints} and \ref{PQT_ES} which show that the matrix elements in the ES-EBCM depend on $R$ as follows:
\begin{align}
\tilde{T}_{\bar{n}\bar{k}} \propto R^{-(n+k+1)}, \nonumber\\[0.2cm]
\tilde{P}_{\bar{n}\bar{k}} \propto R^{-(n+k+1)}, \nonumber\\[0.2cm]
\tilde{Q}_{\bar{n}\bar{k}} \propto R^{(n-k)}. 
\end{align}


\subsection{Axisymmetric particles}

Similarly to the EBCM, the surface integrals $\b L^{11}$ and $\b L^{31}$ simplify substantially for particles with symmetry of rotation (around the $z$-axis).
One major simplification is that only integrals with the same projected angular momentum ($m$) are non-zero, since the integrals over $\phi$ are zero unless $m=m'$. $m$ can then be taken as a fixed parameter (and is sometimes omitted from the notation for simplicity).
The surface integrals defining $\tilde{\b L}^{11}$ and $\tilde{\b L}^{31}$ reduce to integrals over $\theta$ and are all real, which implies that $\tilde{\b L}^{11}$ is symmetric: $\tilde{L}^{11}_{\bar{n}\bar{k}}=\tilde{L}^{11}_{\bar{k}\bar{n}}$.
The matrix elements in $\b P$, $\b Q$, $\b T$, are then also real if $\epsr$ is real.
In the EM case, the $\b T^{22}$ matrix is symmetric by virtue of optical reciprocity \cite{2002Mishchenko}. From the link between the ECBM and ES-EBCM (Eq. \ref{EBCMvsESBCM}) we deduce that the matrix $\tilde{\b T}$ is also symmetric (even though it is not obvious from Eq. \ref{PQT_ES}).

The integral expression can be further simplified by describing the particle surface in spherical coordinates by an equation of the form $r(\theta)$.
Denoting $r_\theta=\d r/\d\theta$, the surface normal and elementary area are
\begin{align}
\b{n}\d S=\left[r\mathbf{e}_r-r_\theta\mathbf{e}_\theta\right]r\sin\theta\d\theta\d\phi
\end{align}
from which we deduce (using $\frac{\partial}{\partial n}=\b n \cdot \nabla$)
\begin{align}
\frac{\pd \Psi^{(1)}_{km}}{\pd n}\d S &= \left[r\frac{\pd \Psi_{km}^{(1)}}{\pd r}-\frac{r_\theta}{r}\frac{\pd \Psi_{km}^{(1)}}{\pd \theta} \right] 
r\sin\theta \d\theta \d\phi.
\end{align}
In line with the notation of Refs. \cite{2002Mishchenko,2012SomervilleJQSRT}, we write\footnote{This expression assumes that
$P_n^m(\cos\theta)$ is defined with the Condon-Shortley phase $(-1)^m$.}:
\begin{align}
d_n^m(\theta) =& (-)^m \sqrt{4\pi}A_n^mP_n^m(\cos\theta),\\[0.2cm]
\tau_n^m(\theta) =& \frac{\d}{\d\theta} d_n^m(\theta).
\end{align}
where $(-)^m$ is shorthand for $(-1)^m$. We can then deduce after a few manipulations:
\begin{align}
\tilde{L}^{11,m}_{nk}=&\frac{1}{2} \int_0^\pi \d\theta \sin\theta 
\left(\frac{r}{R}\right)^{n+k+1}d_n^m \left[k  d_k^m - \frac{r_\theta}{r}\tau_k^m\right] \nonumber\\[0.2cm]
\tilde{L}^{31,m}_{nk}=&\frac{1}{2} \int_0^\pi \d\theta \sin\theta 
\left(\frac{r}{R}\right)^{k-n}d_n^m \left[k  d_k^m - \frac{r_\theta}{r}\tau_k^m\right]\label{Lints}
\end{align}
Note that the symmetry of $\tilde{L}^{11}_{nk}$ can be made more obvious using integrations by part to obtain:
\begin{align}
\tilde{L}^{11,m}_{nk}=\frac{1}{2} \int_0^\pi & \d\theta \sin\theta 
\left(\frac{r}{R}\right)^{n+k+1} \nonumber \\
&\times \frac{\left(\frac{m^2}{\sin^2\theta}+nk\right)d_n^m d_k^m + \tau_n^m\tau_k^m}{n+k+1}.
\end{align}
Note also that all matrices can be computed for $m\ge 0$ only, and the values for $m<0$ derive from
\begin{align}
\tilde{T}_{nk}^{-m} = \tilde{T}_{nk}^{m} 
\end{align}
and identical relations for the others matrices.

\section{Spheroidal particles in the ES-EBCM formalism}
\label{SecSpheroids}

As highlighted in Refs. \cite{2012SomervilleJQSRT,2013SomervilleJQSRT,2015SomervilleJQSRT}, spheroidal particles are a special case
in EM scattering in the context of the $T$-matrix method and this is also the case in electrostatics.
The electrostatics case is even more special as there exists a full analytical solution
of the problem with the separation of variable method. This provides a means to find analytic expressions for the entire $\b T$, $\b P$, and $\b Q$ matrices (see sec. \ref{SecAnalytic}).
For clarity, we focus the discussion in this section on the simple case of a spheroidal dielectric particle in a uniform external field. The general case will be presented in the next section.

We consider a prolate spheroid of semi-axes $a$ along $x,y$ and $c$ along $z$ ($c>a$). It has half-focal length $f=\sqrt{c^2-a^2}$ and eccentricity $e=f/c$.
The surface is then defined in spherical coordinates as:
\begin{align}
r(\theta) = \frac{a}{\sqrt{1-e^2 \cos^2\theta}}.
\label{Eqnrtheta}
\end{align}

\subsection{Special properties}

Because of the reflection symmetry with respect to the $z=0$ plane ($\theta \rightarrow \pi-\theta$),
half of the integrals are zero, namely:
\begin{align}
\tilde{L}^{11,m}_{nk} = \tilde{L}^{31,m}_{nk} = \tilde{P}^{m}_{nk} = \tilde{Q}^{m}_{nk}=\tilde{T}^{m}_{nk}=0 \nonumber\\
\mathrm{~if~} n+k \mathrm{~is~odd}.
\end{align}

Moreover, it was shown in the EM case \cite{2012SomervilleJQSRT} that the dominant terms in some of the other integrals (with $n+k$ even) vanish for spheroids, causing serious numerical issues. In electrostatics, the situation is much simpler, as the corresponding integrals vanish completely \cite{2012SomervilleJQSRT,2014Farafonov}.
As a result, $\tilde{\b Q}^m$ is upper triangular for spheroids: 
\begin{align}
\tilde{Q}^m_{nk}=\tilde{L}^{31,m}_{nk}=0\quad\mathrm{if}\quad n>k,
\end{align}
which can be deduced from the identities proved in Ref. \cite{2012SomervilleJQSRT}.
One immediate consequence is that for a uniform external field (which has $n=1$ only in its series expansion), the inside field is also uniform (it also has $n=1$ only in its expansion since $\tilde{\b Q}^{-1}$ must also be upper triangular).
This is related to the Eshelby conjecture~\cite{2008Liu,2009Kang}, which states that ellipsoids (of which spheroids are a special case) are the only geometry for which the inside field is uniform when placed in an external uniform field, so we may also conjecture that ellipsoids are the only geometry for which $\tilde{\b Q}$  is upper triangular.

\subsection{Integral approach for a uniform field}

Following the standard EBCM approach, we could calculate the matrix $\tilde{\b T}$ from the surface integrals in Eqs. \ref{Lints} and the expressions in Eqs. \ref{PQT_ES}. This method is however cumbersome so we apply it only for illustration to the elements $\tilde{T}_{n1}^{m=0}$, which are sufficient to solve the problem for a uniform external field along the $z$-axis $\b E=E_0\b{\hat{z}}$. The expansion of the external potential has only one term: $\tilde{b}_1^0=-\sqrt{4\pi}RE_0$.
Since $\tilde{\b Q}$ is upper triangular, $\tilde{\b R}=\tilde{\b Q}^{-1}$ is also upper triangular and $\tilde R_{11}^0=\left(\tilde Q_{11}^0\right)^{-1}$, consequently
\begin{align}
\tilde{d}_1^0=\frac{\tilde{b}_1^0}{1+(\epsr-1)\tilde{L}_{11}^{31,m=0}},
\end{align}
and $\tilde{d}_{n}^0=0$ for $n \geq 2$ (the internal field is uniform, as previously discussed).
Moreover,
\begin{align}
\tilde{T}_{n1}^0  = - \tilde{P}_{n1}^0 \left(\tilde Q_{11}^0\right)^{-1} = - \frac{(\epsr-1)\tilde{L}_{n1}^{11,m=0}}{1+(\epsr-1)\tilde{L}_{11}^{31,m=0}},
\end{align}
and the ``scattered'' field expansion coefficients are given by $\tilde{q}_n^0=\tilde{T}_{n1}^0 \tilde{b}_1^0$.

We must now calculate the two $L$-integrals. Firstly:
\begin{align}
\tilde{L}_{11}^{31,m=0} = \frac{a^2}{2 c^2}\int_0^\pi d\theta\sin\theta\frac{\cos^2\theta}{1-e^2\cos^2\theta}.
\end{align}
This is the standard depolarization factor along the long axis of a prolate spheroid, $L_z$ \cite{1941Stratton,1983Bohren}, which and can be expressed explicitly as
\begin{align}
\tilde{L}_{11}^{31,m=0} = L_z=\frac{1-e^2}{e^2}\left[\frac{1}{2e}\ln\frac{1+e}{1-e}-1\right].
\label{EqnLz}
\end{align}
The second integrals, $\tilde{L}_{n1}^{11,m=0}$, can be simplified to:
\begin{equation}
\tilde{L}_{n1}^{11,m=0}=\int_0^\pi\! \d\theta\sin\theta \cos\theta ~\frac{r(\theta)^{n+4}}{2c^2R^{n+2}} P_n(\cos\theta).
\end{equation}
Because of the reflection symmetry, this integrates to zero for $n$ even. For $n$ odd, accurate numerical evaluation of these integrals is difficult because of oscillations and cancellations. However it is possible to find a very simple analytic solution:
\begin{equation}
\tilde{L}_{n1}^{11,m=0}=\frac{1}{n+2}\frac{\textstyle  a^2c f^{n-1}}{R^{n+2}} \quad (n\mathrm{~odd}).
\label{EqnAppA}
\end{equation}
This result is derived in \ref{AppProofInt} using the integral expression and will also result from the derivation
presented in the next section.
We deduce that:
\begin{align}
\tilde{T}_{n1}^{m=0} = 
\begin{cases}
\mathlarger{ \frac{-(\epsr-1)}{1+(\epsr-1)L_z} ~ \frac{1}{n+2}~\frac{a^2c f^{n-1}}{R^{n+2}} } & n\text{ odd}\\[0.4cm]
0 & n \text{ even}
\end{cases}
\label{EqnTn1}
\end{align}

The scattered potential expansion is then (note that it is $R$-independent as expected):
\begin{equation}
\pot_\mathrm{sca} = \sum_n \tilde{q}_n^0 \Psi_{n0}^{(3)} = \frac{\alpha_{zz}E_0}{4\pi\epsilon_0\epsilon_1}\sum_{n\mathrm{~odd}}^\infty\frac{3f^{n-1}}{n+2} \frac{P_n(\cos\theta)}{r^{n+1}},
\label{EquPhiSca}
\end{equation}
where we have introduced the standard dipolar polarizability $\alpha_{zz}$ of the prolate spheroid along its long axis \cite{1941Stratton,1983Bohren,2009Book}:
\begin{equation}
\alpha_{zz} = 4\pi\epsilon_0\epsilon_1 \frac{a^2c}{3}\frac{\epsr-1}{1+(\epsr-1)L_z}.
\end{equation}

\subsection{Revisiting the Rayleigh hypothesis}

Thanks to this simple analytic expression (Eq. \ref{EquPhiSca}), it is straightforward to infer that the series diverges for $r<f$
and converges for $r>f$. This boundary marks the range of validity of the Rayleigh hypothesis for a prolate spheroid in the ES-EBCM framework.
It is the same boundary as found using asymptotic expressions for the coefficients \cite{2014Farafonov,2016Farafonov}, and as found in full-wave EBCM calculations \cite{2016AuguieJO}. This is expected given the link between the ES-EBCM and the long-wavelength limit of the EBCM. The advantage of our analytic expression is that it provides a vivid illustration of the mathematical concepts involved in discussing the Rayleigh hypothesis, in particular its link to the analytic continuation of the solution, on which we now elaborate.

If we consider a high aspect ratio spheroid with $c/a>\sqrt{2}$, there is a region close to the surface (away from the tips), where $r<f$ and the scattered field expansion diverges. Despite this, the ES-EBCM (or EBCM) remains valid as it does not rely on the Rayleigh hypothesis and the expansion in Eq.~\ref{EquPhiSca} remains 
correct where it converges -- it only fails in some regions near the surface. The reason for this failure is that the spherical basis chosen to expand the scattered field is not well suited to the intrinsic singularity of the solution.
To see this explicitly, one can find an analytic expression for the series expansion in Eq.~\ref{EquPhiSca} (derived in \ref{AppAnalCont}):
\begin{equation}
\pot_\mathrm{sca} = \frac{3\alpha_{zz}E_0}{4\pi\epsilon_0 \epsilon_1 f^3}  \left[ z\ln\left(\frac{r'-z'}{r''-z''}\right) + r'-r''\right], 
\label{EquAnalCont}
\end{equation}
where $r',z'$ and $r'',z''$ refer to the coordinates in the offset frames centered on the focal points F' (at ${\b r}=f\hat{\b z}$) and F''
(at ${\b r}=-f\hat{\b z}$), or explicitly:
\begin{align}
z' &=z-f, ~\quad r' =\sqrt{r^2-2fr\cos\theta+f^2},&\nonumber\\
z''&=z+f, ~\quad r''=\sqrt{r^2-2fr\cos\theta+f^2}.& \label{offset_frames}
\end{align}
This is the analytic continuation of Eq.~\ref{EquPhiSca}; it is singular only on the $z$ axis between the two foci of the spheroid.
The basis of solid spherical harmonics was not well suited for such a solution because a series of these harmonics always has a spherical boundary of convergence, and such an expansion (Eq.~\ref{EquPhiSca}) therefore increased the singular region from a line to the smallest sphere containing that line (the sphere $r=f$).
Note nevertheless that since the analytic continuation of a series is unique, the series in Eq.~\ref{EquPhiSca} already contains, at least formally, all the information about the scattered field, even in the regions where it diverges.
In general however, finding an analytic continuation is a difficult problem, and other methods must be used to calculate the scattered field near the surface \cite{2016AuguieJO}. These arguments have been laid out before from an abstract point of view \cite{1969BurrowsEL1,1969BatesEL,1969BurrowsEL2,1969MillarEL,1975Bates,1985Maystre,2009Rother}, but we believe the simple analytic expressions derived here provide an insightful illustration.

\section{Full analytic solution of the ES-EBCM formalism for prolate spheroids}
\label{SecAnalytic}

In the case of the electrostatics problem for spheroids, the general solution may also be found using separation of variables in spheroidal coordinates. Those solutions can then be expressed in the basis of solid spherical harmonics to find analytic expressions for all the matrix elements of the ES-EBCM.

\subsection{Separation of variables method}

We use prolate spheroidal coordinates $(\xi,\eta,\phi)$ defined as follows~\cite{1953Morse}: 
\begin{align}
\xi=\frac{r''+r'}{2f}, \qquad \eta=\frac{r''-r'}{2f},
\label{EqnProlCoord}
\end{align}
where $r'$, $r''$ were defined in Eq.~\ref{offset_frames} and $\phi$ is the same as in spherical coordinates. Surfaces of constant $\xi$ represent prolate spheroids with focal length $2f$.
We will consider a dielectric prolate spheroid of semi-axes $a$ and $c$ ($c>a$) as in the previous section in a  general external field.
The half focal length is $f=\sqrt{c^2-a^2}$ and the eccentricity $e=f/c$. In spheroidal coordinates, the surface is at $\xi=\xi_0=1/e$.

For the most general external potential, this problem can be solved using solid spheroidal harmonic bases by expanding the potentials as:
\begin{align}
\pot_\mathrm{ext} &= \sum_{n,m} \alpha_n^m P_n^m(\xi)P_n^m(\eta)e^{im\phi}, \label{EqnPhiExtProl}\\
\pot_\mathrm{in}  &= \sum_{n,m} \beta_n^m P_n^m(\xi)P_n^m(\eta)e^{im\phi}, \\
\pot_\mathrm{sca}  &= \sum_{n,m} \gamma_n^m Q_n^m(\xi)P_n^m(\eta)e^{im\phi}, \label{EqnPhiScaProl}
\end{align}
where $Q_n^m$ are the associated Legendre functions of the second kind.

The boundary conditions are:
\begin{align}
\pot_\mathrm{in}|_{\xi=\xi_0}=\pot_\mathrm{out}|_{\xi=\xi_0}; & \qquad \epsr \left.\frac{\pd \pot_\mathrm{in}}{\pd \xi}\right|_{\xi=\xi_0} = \left.\frac{\pd \pot_\mathrm{out}}{\pd \xi}\right|_{\xi=\xi_0} .
\end{align}
Applying these and using the fact that $P_n^m(\eta)$ form a basis, we have
\begin{align}
\beta_n^m=\alpha_n^m+\gamma_n^m\frac{Q_n^m(\xi_0)}{P_n^m(\xi_0)}, \quad
\epsr \beta_n^m=\alpha_n^m+\gamma_n^m\frac{\Qnmp(\xi_0)}{\Pnmp(\xi_0)},
\end{align}
where $\Pnmp$ and $\Qnmp$ denote the derivatives of $P_n^m$ and $Q_n^m$.
From this, we derive
\begin{align}
\beta_n^m=\Gamma_n^m\alpha_n^m; \quad \gamma_n^m=\Upsilon_n^m\alpha_n^m, \label{gam_vs_alph}
\end{align}
where
\begin{align}
\Gamma_n^m &=\frac{\Qnmp(\xi_0)P_n^m(\xi_0)-Q_n^m(\xi_0)\Pnmp(\xi_0)}{\Qnmp(\xi_0)P_n^m(\xi_0) -\epsr Q_n^m(\xi_0)\Pnmp(\xi_0)}, \nonumber \\[0.2cm]
\Upsilon_n^m &=\frac{(\epsr-1)P_n^m(\xi_0)\Pnmp(\xi_0)}{\Qnmp(\xi_0)P_n^m(\xi_0) -\epsr Q_n^m(\xi_0)\Pnmp(\xi_0)}.
\label{EqnSuscep1}
\end{align}
We have introduced the external $\Upsilon_n^m$ and internal $\Gamma_n^m$ spheroidal susceptibilities, which are similar to the multipolar polarizabilities for a sphere but depend on $m$, as expected for a non-spherically-symmetric object.
$\Upsilon_n^m$ is analogous to the $\tilde{\b T}$ matrix: for a spheroid, $\tilde{\b T}^m$ is diagonal in the spheroidal basis with diagonal elements $\Upsilon_n^m$. Similar observations can be made for the other matrices: $\tilde{\b P}$, $\tilde{\b Q}$, $\tilde{\b L}^{11}$,
$\tilde{\b L}^{31}$. For example $(\Gamma_n^m)^{-1}$ are the diagonal elements of $\tilde{\b Q}^m$ and $-\Upsilon_n^m/\Gamma_n^m $ the diagonal elements of $\tilde{\b P}$ in the spheroidal basis.

The spheroidal susceptibilities can be simplified by identifying the Wronskian of the associated Legendre equation (see Sec.~8.1.8 in Ref.~\cite{Abramovitz}):
\begin{align}
W_n^m(\xi_0) &= P_n^m(\xi_0)\Qnmp(\xi_0) - \Pnmp(\xi_0)Q_n^m(\xi_0)	\nonumber\\
&=\frac{(-)^{m+1}}{\xi_0^2-1}\frac{(n+m)!}{(n-m)!} =(-)^{m+1}\frac{(n+m)!}{(n-m)!}\frac{f^2}{a^2}. \label{Wronsk}
\end{align}
Then we have 
\begin{align}	
\Gamma_n^m &= \frac{1}{1 - (\epsr-1) Q_n^m(\xi_0)\Pnmp(\xi_0)/W_n^m(\xi_0)},	\\[0.2cm]
\Upsilon_n^m &= \frac{(\epsr-1)P_n^m(\xi_0)\Pnmp(\xi_0)}{W_n^m(\xi_0)-(\epsr-1) Q_n^m(\xi_0)\Pnmp(\xi_0)},   \\[0.2cm]	
\frac{\Upsilon_n^m}{\Gamma_n^m} &= (\epsr-1)\frac{P_n^m(\xi_0)\Pnmp(\xi_0)}{W_n^m(\xi_0)}. 
\end{align}
We can also rearrange the solution as:
\begin{align}
\gamma_n^m &= -(\epsr-1) \Lambda_n^{11,m}  \beta_n^m \label{EqnL11Prol}\\[0.2cm]
\alpha_n^m-\beta_n^m &= (\epsr-1) \Lambda_n^{31,m} \beta_n^m, \label{EqnL31Prol}
\end{align}
where
\begin{align}
\Lambda_n^{11,m} &= -\frac{P_n^m(\xi_0) \Pnmp(\xi_0)}{W_n^m(\xi_0)},\\[0.2cm]
\Lambda_n^{31,m} &= -\frac{Q_n^m(\xi_0) \Pnmp(\xi_0)}{W_n^m(\xi_0)}, 
\end{align}
are the diagonal elements of $\tilde{\b L}^{11}$, $\tilde{\b L}^{31}$ expressed in the spheroidal basis.

\subsection{Analytic expressions for all matrix elements}

Using the simple relationships between the incident, scattered and internal fields in the spheroidal basis, and the expansions between spherical and spheroidal harmonics, we can derive the equivalent relationships in the spherical basis and therefore analytic expressions for all elements of $\tilde{\b P}$, $\tilde{\b Q}$, and $\tilde{\b T}$.
The spherical-spheroidal harmonic expansions \cite{2000Jansen,2002Russian} are summarized in \ref{AppExpansions}. These express the spherical basis functions, $r^nP_n^m(\cos\theta)$ or $r^{-n-1}P_n^m(\cos\theta)$, as a sum of spheroidal basis functions, $P_k^m(\xi)P_k^m(\eta)$ or $Q_k^m(\xi)P_k^m(\eta)$, and vice versa. We set $R=f$ in the following to simplify the resulting expressions.

Our goal is to express $\tilde{q}_n^m$ in terms of $\tilde{b}_n^m$.
We first express $\alpha_n^m$ in terms of $\tilde{b}_n^m$, i.e. derive the spheroidal expansion of the external field from its spherical expansion.
Inserting Eq.~\ref{PvsPP} into the external potential spherical expansion (Eq.~\ref{EqnPhiExtSph}), rearranging the order of summation, and identifying the coefficients $\alpha_n^m$ from Eq.~\ref{EqnPhiExtProl}, we obtain:
\begin{align}
\alpha_s^m&= \sum_{k=s}^\infty{\scriptstyle\underline{sk}}~ \frac{(2s+1)(k+m)!}{(k-s)!!(k+s+1)!!}\frac{(s-m)!}{(s+m)!} A_k^m \tilde{b}_k^m, \label{alph_vs_b} \\ 
&\text{with }\quad {\scriptstyle\underline{sk}} =
\begin{cases} 1 & \text{if $s+k$ even,} \\
0 &\text{if } s+k \text{ odd.} \end{cases}
\end{align}
We then express $\tilde{q}_n^m$ in terms of $\gamma_n^m$ by substituting Eq.~\ref{PvsQP} into Eq.~\ref{EqnPhiScaProl} and identifying the coefficients $\tilde{q}_n^m$ from Eq.~\ref{EqnPhiScaSph}
\begin{align}
\tilde{q}_n^m &= \sum_{s=|m|}^n{\scriptstyle\underline{sn}}~ \frac{(-)^m(n-m)!}{(n-s)!!(s+n+1)!!}\frac{(s+m)!}{(s-m)!} \frac{1}{A_n^m} \gamma_s^m  \label{q_vs_gam} 
\end{align}
The relationship between $\tilde{q}_n^m$ and $\tilde{b}_n^m$ is found by combining Eqs. \ref{alph_vs_b}, \ref{q_vs_gam} with the relation 
$\gamma_s^m=\alpha_s^m\Upsilon_s^m$ (Eq. \ref{gam_vs_alph}) and swapping the order of summation:
\begin{align}
\tilde{q}_n^m &=\sum_{k=|m|}^\infty\sum_{s=|m|}^{\min(n,k)}{\scriptstyle\underline{sk}}~{\scriptstyle\underline{sn}}\frac{A_k^m}{A_n^m}\Upsilon_s^m\nonumber\\
& \times\frac{(-)^m(2s+1)(n-m)!(k+m)!}{(n-s)!!(n+s+1)!!(k-s)!!(k+s+1)!!}    \tilde{b}_k^m,
\end{align}
where $\min(n,k)$ denotes the minimum of $n$ and $k$.
The matrix elements of $\mathbf{\tilde{T}}$ can be readily identified as:
\begin{align}	
	&\tilde{T}_{nk}^m ={\scriptstyle\underline{nk}}(-)^m C_n^m C_k^m \times  \label{EqnTanal} \\
	&\sum_{s=|m|}^{\min(n,k)} 	\frac{{\scriptstyle\underline{sn}} (2s+1)~ \Upsilon_s^m}{(n-s)!!(n+s+1)!!(k-s)!!(k+s+1)!!} , \nonumber 
\end{align}
where we have defined
\begin{align}
C_n^m = \sqrt{(n-m)!(n+m)!}.
\end{align}
Note that $\Upsilon_s^m$ depends on the spheroid shape (via $\xi_0$) and its optical properties (via $\epsilon$).

For $\tilde{\b L}^{11}$, we use Eqs. \ref{EqnL11Prol} and the derivation is almost identical to that of $\tilde{\b T}$:
\begin{align}	
	&\tilde{L}^{11,m}_{nk} ={\scriptstyle\underline{kn}}(-)^m C_n^m C_k^m \times \label{EqnL11anal}\\
	& \sum_{s=|m|}^{\min(n,k)} 
	\frac{{\scriptstyle\underline{sn}}(2s+1)~\Lambda_s^{11,m}}{(n-s)!!(n+s+1)!!(k-s)!!(k+s+1)!!} . \nonumber
\end{align}
In this form, the symmetry of $\tilde{\b{T}}$ and $\tilde{\b{L}}^{11}$ is clear.
For $\tilde{\b L}^{31}$, we need the inverse relation of Eq.~\ref{alph_vs_b}, which is obtained using
the expansion Eq.~\ref{PPvsP}. We then use Eq.~\ref{EqnL31Prol} to get:
\begin{align}
	&\tilde{L}^{31,m}_{nk} = 	0  \quad  \mathrm{if~}n>k,\quad\mathrm{else} \nonumber \\
	&\tilde{L}^{31,m}_{nk} = 	{\scriptstyle\underline{kn}}\frac{C_k^m}{C_n^m} \times \label{EqnL31anal}\\
&\sum_{s=n}^k {\scriptstyle\underline{sn}} \frac{(-)^{(s-n)/2}(2s+1)(n+s-1)!!~\Lambda_s^{31,m}}{(s-n)!!(k-s)!!(k+s+1)!!}\nonumber
\end{align}
As previously observed, the $L$-matrices only depend on shape, not on $\epsr$. 
$\tilde{\b{P}}$ and $\tilde{\b{Q}}$ are easily derived from Eqs. \ref{PQT_ES}.

\subsection{Special cases}

It is interesting to consider a few special cases.
Firstly for $m=0$:
\begin{align}
\Lambda^{31,m=0}_{n=1} = (\xi_0^2-1)(\xi_0 Q_0(\xi_0)-1) = L_z.
\end{align}
The latter equality arises when setting $e=1/\xi_0$ in Eq. \ref{EqnLz}.
We also have
\begin{align}
\Lambda^{11,m=0}_{n=1} = (\xi_0^2-1)\xi_0  = \frac{1-e^2}{e^3}=\frac{a^2 c}{f^3}.
\end{align}
From those we deduce
\begin{align}
\Upsilon^{0}_1 = -\frac{(\epsr-1)}{1+(\epsr-1)L_z} \frac{a^2 c}{f^3}.
\end{align}

For the $T$-matrix, we have from Eq. \ref{EqnTanal}:
\begin{align}
\tilde{T}_{n1}^{m=0} = \frac{\Upsilon_1^0}{n+2} \qquad (n\text{ odd}),
\end{align}
which is the same as Eq.~\ref{EqnTn1} found earlier.

For $m=1$, we have
\begin{align}
\Lambda^{11,m=1}_{n=1} &= -\frac{a^2 c}{2f^3},\\[0.2cm]
\Lambda^{31,m=1}_{n=1} &= \frac{1}{2}\left[\frac{c^2}{f^2} - \xi_0 Q_0(\xi_0)\right] = \frac{1- L_z}{2}.
\end{align}
The latter is the depolarization factor of the spheroid along the $x,y$ directions: $L_x= L_y=(1-L_z)/2$.
From those, we have
\begin{align}
\Upsilon^{1}_1 = \frac{(\epsr-1)}{1+(\epsr-1)L_x} \frac{a^2 c}{2f^3},
\end{align}
and Eq. \ref{EqnTanal} simplifies to
\begin{align}
\tilde{T}_{n1}^{m=1} = -\sqrt{\frac{2(n+1)}{n}}\frac{\Upsilon_1^1}{n+2}.
\end{align}

Finally, for a general $n$ and $m$, Eq.~\ref{EqnL31anal} also reduces to a single term when $k=n$:
\begin{align}
\tilde{L}^{31,m}_{nn}=\Lambda^{31,m}_n = -\frac{Q_n^m(\xi_0) \Pnmp(\xi_0)}{W_n^m(\xi_0)}.
\end{align}
Those diagonal elements of $\tilde{L}^{31}$ can be viewed as the generalized depolarization
factors of the spheroid for multipole orders $n,m$.

\begin{figure*}
\includegraphics[width=\textwidth,trim={0cm 6cm 0cm 5cm},clip]{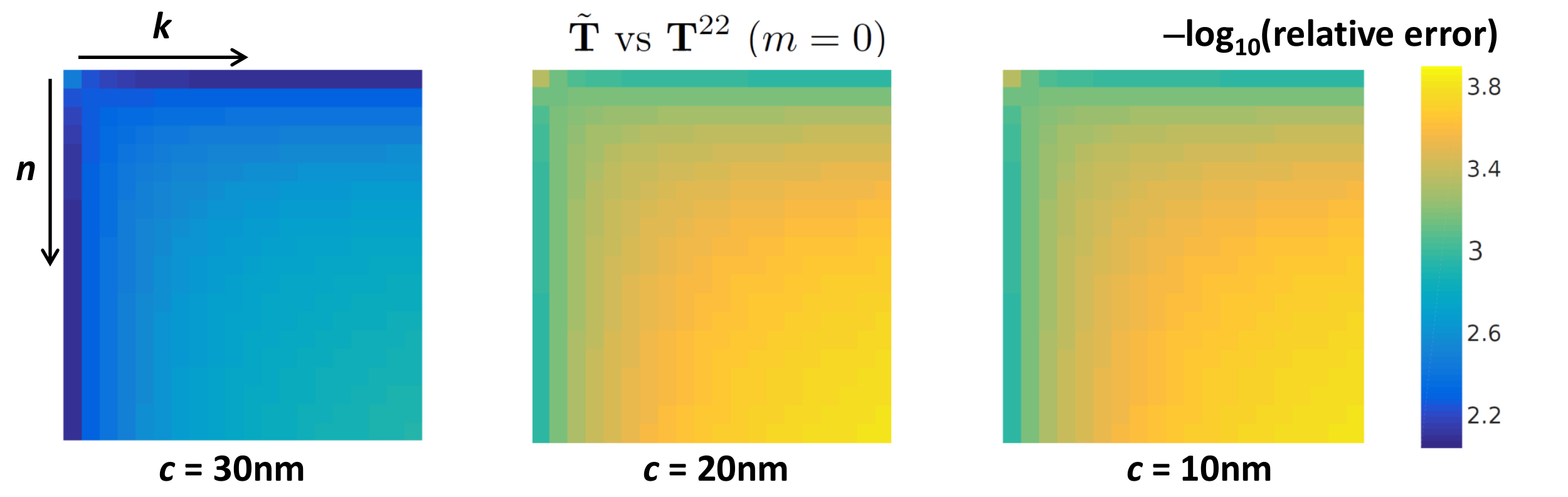} 
\caption{
Colormaps of the number of digits of agreement (computed from $-\log_{10}$ of the relative error) between two methods of calculating the $T$-matrix for prolate spheroids of long-axis $c$: using the EBCM code from \cite{2016SMARTIES} and from the electrostatic analytic formula given in Eq. \ref{EqnTanal}. In each plot, the aspect ratio is $c/a$=10, the wavelength of excitation is 600\,nm and the relative permittivity is $\epsilon=1.5$. The EBCM $T$-matrix was compared to that obtained from the ES-EBCM using Eq.~\ref{EBCMvsESBCM}. The elements shown here are for $m=0$ (similar results are obtained for higher $m$), and $n,k\leq39$ (only the non-zero elements are shown). The lower degree elements are towards the top left. The error decreases with particle size as the problem approaches the quasi-static limit.} \label{fig}
\end{figure*}

\subsection{Comparison with EBCM results} 

To assess the applicability of our analytic results, we compare the analytic matrix elements for spheroids to those obtained numerically using an accurate implementation of the EBCM method for spheroids \cite{2015SomervilleJQSRT} for which codes are available \cite{2016SMARTIES}. The relations established in Eq.~\ref{EBCMvsESBCM} are used to compare the two methods. For particles much smaller than the wavelength, the analytic quasi-static solution is expected to provide a good approximation to the exact result. This is illustrated in Fig. \ref{fig} for three dielectric prolate spheroids all of aspect ratio 10 and long semi-axis 10, 20 and 30\,nm, excited with 600\,nm wavelength light. 
The agreement between the two methods increases as the size decreases, as expected.
Note however that the other blocks of the EBCM $T$-matrix, $\b T^{11},~\b T^{12}$, and $\b T^{21}$ contain elements that are not negligible in the quasi-static limit. So for the ES-EBCM to be applicable as an approximation to quasi-static (not just electrostatic) problems, these other blocks should also be taken into account.

\section{Conclusion}
\label{SecConclusion}

In summary, this work provides a detailed investigation of the electrostatics limit of Waterman's $T$-matrix formalism, in particular in the special case of
spheroidal particles. We focused specifically on prolate spheroids but the results could be adapted to oblate spheroids.
We believe that the analytic expressions that we derived will provide an insight into the fundamental properties of the $T$-matrix formalism. This was illustrated in the context of the Rayleigh Hypothesis and its connection to the analytic continuation of the solutions.
The ES-EBCM formalism could be further developed to encompass the magnetostatic limit (relating to magnetic multipoles) and the long-wavelength limit of the coupling matrices between electric and magnetic multipoles.
The ES-EBCM formalism could moreover be applied to the solution of other electrostatic problems.

\appendix

\section{Proof of Eq.~\ref{EqnAppA}}
\label{AppProofInt}

We reproduce here the form of the integral
\begin{equation}
\tilde{L}_{n1}^{11,m=0}=\int_0^\pi \d\theta\sin\theta \cos\theta ~\frac{r(\theta)^{n+4}}{2c^2R^{n+2}} P_n(\cos\theta)
\end{equation}
and consider $n$ odd only (the integral is zero for $n$ even). 
We first substitute $r(\theta)$ from Eq. \ref{Eqnrtheta} and make the change of variable $x=\cos\theta$:
\begin{equation}
\tilde{L}_{n1}^{11,m=0}=\frac{a^{n+4}}{2c^2R^{n+2}}\!\int_{-1}^{1} \d x\! \frac{x}{\left(1-e^2 x^2\right)^{\frac{n+4}{2}}} P_n(x).
\end{equation}
Using the binomial expansion, this is rewritten as a series
\begin{equation}
\tilde{L}_{n1}^{11,m=0}=\frac{a^{n+4}}{2ec^2R^{n+2}}\int_{-1}^{1} \d x \sum_{p=0}^{\infty} \alpha_{p,n}(ex)^{2p+1} P_n(x),
\end{equation}
where
\begin{equation}
\alpha_{p,n}=\binom{p+\frac{n}{2}+1}{p}. 
\end{equation}
Note that $n$ is odd and this is a generalized binomial coefficient with half-integer entries.

We then expand the powers $x^{2p+1}$ on the basis of Legendre polynomials
as
\begin{equation}
x^{2p+1}=\sum_{l\mathrm{~odd}}^{2p+1} (2l+1) \beta_{l,p} P_l(x)
\end{equation}
This is as a special case of Eq. \ref{PvsPP}, evaluated at $\eta=1$, $n=2p+1$, $m=0$, where the coefficients are
\begin{equation}
\beta_{l,p} = \frac{2^{2l}(2p+1)!(p+\frac{l+1}{2})!}{(p-\frac{l-1}{2})!(2p+l+2)!}.
\end{equation}

Using the orthogonality of the Legendre polynomials, the integral can be expressed as
\begin{align}
\tilde{L}_{n1}^{11,m=0}=\frac{a^{n+4}}{c^2R^{n+2}} \sum_{p=(n-1)/2}^{\infty} \alpha_{p,n}\beta_{n,p} e^{2p}.
\label{EqInt1}
\end{align} 
$\alpha_{p,n}\beta_{n,p}$ can be shown to simplify to 
\begin{align}
\alpha_{p,n}\beta_{n,p}=\frac{1}{n+2}\binom{p+\frac{1}{2}}{p-\frac{n-1}{2}}.
\end{align}
Re-indexing the sum with $k=p-(n-1)/2$:
\begin{equation}
\tilde{L}_{n1}^{11,m=0}=\frac{a^{n+4}e^{n-1}}{c^2R^{n+2}(n+2)} \sum_{k=0}^{\infty} \binom{k+\frac{n}{2}}{k}e^{2k},
\end{equation}
where we recognize the binomial series expansion for $\frac{1}{(1-e^2)^{n/2+1}}$, which simplifies
using $e=f/c$ and $1-e^2=a^2/c^2$ to
\begin{equation}
\tilde{L}_{n1}^{11,m=0}=\frac{a^2 c f^{n-1}}{R^{n+2}}\frac{1}{n+2}.
\end{equation}
As shown in Sec.~\ref{SecAnalytic}, this can be alternatively derived from the spheroidal coordinate solution of the problem.
Given the remarkable simplicity of the final expression for $\tilde{L}_{n1}^{11,m=0}$, one might have expected that a simpler proof would be possible.

\section{Analytic continuation of the multipole expansion for the scattered potential}
\label{AppAnalCont}

We start from the series expansion given in the main text (rewritten here up to a constant factor):
\begin{equation}
\pot(r,\theta) = \sum_{n\mathrm{~odd}}^\infty\frac{1}{n+2} \left(\frac{f}{r}\right)^{n+1} P_n(\cos\theta).
\end{equation}

We first recognize that this expression has some resemblance with the generating function for the
Legendre polynomials, namely:
\begin{equation}
\frac{1}{\sqrt{1-2u\cos\theta+u^2}}=\sum_{n=0}^\infty u^n P_n(\cos\theta).
\end{equation}

We therefore define 
\begin{align}
G(u,\theta) &= \frac{1}{u}\int_0^u\! \frac{v\d v}{\sqrt{1-2v\cos\theta+v^2}} \nonumber\\
&= \sum_{n=0}^\infty \frac{u^{n+1}}{n+2} P_n(\cos\theta)
\end{align}
from which we infer
\begin{equation}
\pot(r,\theta) = \frac{1}{2}\left[ G\left(\frac{f}{r},\theta\right)-G\left(-\frac{f}{r},\theta\right)\right].
\end{equation}
The integral for $G$ can be calculated analytically:
\begin{align}
G&(u,\theta) = \frac{\sqrt{1-2u\cos\theta+u^2}-1}{u} \nonumber\\[0.2cm]
&+ \frac{\cos\theta}{u}\ln\left(\frac{u-\cos\theta+\sqrt{1-2u\cos\theta+u^2}}{1-\cos\theta}\right).
\end{align}
Note that this result can also be obtained from Eq. \ref{QPvsP} with $n=0$.
The simpler expression given in Eq. \ref{EquAnalCont} can be obtained by considering the offset coordinate frames centred at both foci (the offset coordinates are defined explicitly in Eqs. \ref{offset_frames}).

\section{Relationships between spherical and spheroidal solid harmonics}
\label{AppExpansions}

Below are the four possible relations between the regular and irregular spherical solid harmonics, and the regular and irregular prolate spheroidal harmonics. The azimuthal dependence $e^{\pm im\phi}$ is omitted since it is the same on both sides. The spheroidal coordinates are defined in Eq. \ref{EqnProlCoord}. By setting $R=f$ in the definitions of our spherical basis functions, those can easily be rewritten in terms of $\Psi_{nm}^{(1)}$ and $\Psi_{nm}^{(3)}$. Derivations can be found in \cite{2000Jansen,2002Russian}.
\begin{align}
&P_n^m(\xi)P_n^m(\eta)=\frac{(n+m)!}{(n-m)!} \sum_{k=m}^n{\scriptstyle\underline{kn}} ~(-)^{(n-k)/2}\nonumber\\
&\quad\times\frac{(n+k-1)!!}{(n-k)!!(k+m)!}\left(\frac{r}{f}\right)^kP_k^m(\cos\theta) \label{PPvsP}
\\[0.6cm]
&\left(\frac{r}{f}\right)^nP_n^m(\cos\theta)=(n+m)!\sum_{k=m}^n{\scriptstyle\underline{kn}} \nonumber\\
&\quad\times\frac{(2k+1)}{(n-k)!!(n+k+1)!!}\frac{(k-m)!}{(k+m)!} P_k^m(\xi)P_k^m(\eta) \label{PvsPP}
\\[0.6cm]
&Q_n^m(\xi)P_n^m(\eta)=(-)^m\frac{(n+m)!}{(n-m)!}\sum_{k=n}^\infty{\scriptstyle\underline{kn}}~\nonumber\\
&\quad\times\frac{(k-m)!}{(k-n)!!(k+n+1)!!}\left(\frac{f}{r}\right)^{k+1}\!P_k^m(\cos\theta) \label{QPvsP}
\\[0.6cm]
&\left(\frac{f}{r}\right)^{n+1}\!P_n^m(\cos\theta)=\frac{(-)^m}{(n-m)!}\sum_{k=n}^\infty{\scriptstyle\underline{kn}} ~(-)^{(n-k)/2}\nonumber\\
&\quad\times\frac{(2k+1)(n+k-1)!!}{(k-n)!!}\frac{(k-m)!}{(k+m)!} Q_k^m(\xi)P_k^m(\eta) \label{PvsQP} 
\end{align}

\section*{Acknowledgments}
We acknowledge the support of the Royal Society of New Zealand (RSNZ) through a Marsden Grant (ECLR)
and a Rutherford Discovery Fellowship (BA).\\
\hrule
\newpage

\bibliographystyle{elsarticle-num}
\bibliography{TmatrixESA_v7}
\end{document}